\documentstyle[12pt]{article}
\newcommand \beq{\begin{eqnarray}}
\newcommand \eeq{\end{eqnarray}}
\newcommand \ga{\raisebox{-.5ex}{$\stackrel{>}{\sim}$}}
\newcommand \la{\raisebox{-.5ex}{$\stackrel{<}{\sim}$}}

\begin{document}
\bibliographystyle{unsrt}
\baselineskip=18pt

\title{\bf Physics of Coulomb Corrections in Hanbury-Brown Twiss
Interferometry in Ultrarelativistic Heavy Ion Collisions
}
\author{
   Gordon Baym\\
   Department of Physics\\
    University of Illinois at Urbana-Champaign,\\
          1110 W. Green St., Urbana, IL 61801, U.S.A.\\
and\\
   Peter Braun-Munzinger\\
   Gesellschaft f\"ur Schwerionenforschung (GSI)\\
   Planckstr. 1, 64220 Darmstadt, GERMANY\\
   }
\date{\today}
\maketitle

\begin{abstract}

    \noindent We discuss the elementary physics of the final state Coulomb
interactions in Hanbury-Brown Twiss interferometry, showing -- with explicit
comparison to E877 data for $\pi^+\pi^-$ and $\pi^\pm p$ -- that the Coulomb
corrections in the pair correlation function can be well understood in terms
of simple classical physics.  We connect the classical picture with
descriptions in terms of Coulomb wave functions, and investigate the influence
of the ``central'' Coulomb potential on the pair correlation function.

\end{abstract}


\section{Introduction}

    Hanbury-Brown Twiss interferometry of identical mesons has become an
important probe of the evolving geometry of the collision volume in
ultrarelativistic heavy-ion collisions [1-3].
The quantitative interpretation of the results depends critically on
understanding the role of Coulomb interactions of the detected pairs of
particles with each other, as well as the Coulomb interactions of the pair
with the system of remaining particles.  In this note we focus on the
elementary physics of these processes, deferring detailed calculations to later
publications.

    The simplest form of Coulomb correction is inclusion of the Gamow factor
-- the square of the relative Coulomb wave function, $\psi_C(0)$, of the
produced pair at zero separation -- a procedure which is followed in many
analyses \cite{gyu79,gyu81}.  The assumption is that the produced pair of
identical particles is made in a relative Coulomb state (at zero separation),
and the amplitude for doing so is thus modified from the bare amplitude by the
factor $\psi_C(0)$; non-relativistically
\beq
   \psi_C(0) = \left(\frac{2\pi\eta}{e^{2\pi\eta}-1}\right)^{1/2},
\label{psi0}
\eeq
where the dimensionless parameter $\eta$ is given by
\beq
  \eta = \frac{zz^\prime e^2}{v_{rel}} = \frac{zz^\prime \alpha }{v_{rel}/c},
\label{eta}
\eeq
for a pair of particles of charges $z e$ and $z^\prime e$ with relative
momentum $\vec q = (\vec p\, - \vec p\,^\prime)/2$ and relative velocity
$v_{rel} = q/m_{red}$, with reduced mass $m_{red} = m/2$ for two particles of
mass $m$.  [We consider particles with $|z|$ = 1 here.] The Coulomb-corrected
rate of production is then inferred to be the measured rate divided by
$|\psi_C(0)|^2$.  For pairs of the same charge, the Gamow correction
suppresses the probability of production at small $q$, by a factor tending at
small $q$ to $2\pi\eta e^{-2\pi\eta}$, while enhancing the probability of
production for opposite sign pairs by a factor tending to $2\pi|\eta|$ at
small $q$.

    Correcting for Coulomb effects by taking the relative Coulomb wave
function at the origin is not physically correct in heavy ion collisions.  As
noted in Refs.~\cite{pratt86,bowler} (see also [8-11]),
taking the finite size of the source size into account can produce
significant effects.  The standard Gamow correction assumes that the
separation of the particles of the pair at creation is small compared with the
(zero angular momentum) {\it classical turning point}, $r_t$, defined by
$q^2/2m_{red} = e^2/r_t$.  However, for pions, $r_t\simeq (200$ fm)/$q^2$,
where $q$ is measured in MeV/c; for $q\sim$ 10 MeV/c, a typical minimum value,
$r_t$ is only 2 fm, and smaller for larger $q$.  Since $r_t$ is much smaller
than the characteristic heavy ion radius, most of the pairs of particles
observed in a heavy ion collision are made at relative separations well
outside their classical turning points.

    We note that for typical $q$, the three length scales in the problem --
the turning point, the particle wavelength, and the two-particle Bohr radius,
$a_0 = 1/m_{red}e^2$ (= 387 fm for $\pi\pi$ and 222 fm for $\pi$p) -- are
cleanly separated:
\beq
 r_t:1/q:a_0 = 2: a_0q: (a_0q)^2.
\label{scales}
\eeq
For $\pi\pi$ (or $\pi$p), $a_0q = 1/|\eta| = 1.96$ (or 1.13) $q/({\rm
Mev/c})\gg 1$.  The classical turning point is the relevant length scale here
for Coulomb effects (not, as suggested in Refs.  \cite{bauer,pratt86}, the
two-particle Bohr radius).

    Furthermore, in the presence of many produced particles, the relative
Coulomb interaction of a pair is highly screened, which also decreases effects
of Coulomb suppression.  The motion of the particles in the pair is strongly
affected by their interactions with the plasma of other particles, and the
mutual Coulomb interaction of the pair becomes dominant only when the pair has
sufficiently separated from the other particles in the system that there is
small probability of finding other particles between the particles in the
pair.  (See Ref.  \cite{anchishkin} for a recent investigation of screening
effects in the context of a particular model for high particle
multiplicities.)

    The major effect of the Coulomb interaction between the particles of the
pair, at distances large compared with $r_t$, is to accelerate them relative
to each other.  Particles of the same charge are accelerated to larger
relative momenta, thus depressing the observed distribution at small $q$,
while particles of opposite charge are reduced in relative momentum in the
final state, which builds up the distribution at small $q$.  Although these
effects are qualitatively similar those produced by the Gamow correction, they
are quantitatively rather different.

    Our main focus in this paper is on a simple schematic model for effects of
Coulomb interactions.  As we shall see, correcting for the Coulomb final state
interaction of non-identical pairs enables one to extract important
information contained in the measured correlation function, about the spatial
and temporal size of the emitting source. The same detailed information
about the Coulomb final state interaction is also important to correct
correlation functions of identical particles for Coulomb effects.

\section{Toy model}

    We construct a greatly simplified model to take the screening and
acceleration effects into account by neglecting the Coulomb interaction
between the pair for separations less than an initial radius $r_0$, and for
separations greater than $r_0$ including only the relative Coulomb
interaction.  Since the relative motion is in the classical region,
conservation of energy of the pair implies that the final observed relative
momentum $q$ is related to the initial momentum of the pair $q_0$ at $r_0$ by
(see, e.g., \cite{gyu81,koonin79}
\beq \frac{q^2}{2m_{red}} =
\frac{q_0^2}{2m_{red}} \pm \frac{e^2}{r_0},
\label{toy}
\eeq
where the upper sign is for particles of like charge, and the lower for
particles of opposite charge.  For example, for pions with $r_0$ = 10 fm, $q^2
= q_0^2 \pm 20(\rm{MeV/c})^2$.

    How does the acceleration affect the measured correlation function?  In a
heavy ion collision the distribution of singles is given in terms of particle
creation and annihilation operators by
\beq
n(\vec p) =
      \langle a^\dagger(\vec p\,) a(\vec p\,)\rangle,
\label{singles}
\eeq
while the distribution of pairs of particles of momenta $\vec p\,$ and
$\vec p\,^\prime$ is given by
\beq
n_2(\vec p\,,\vec p\,^\prime) =
      \langle a^\dagger(\vec p\,) a^\dagger(\vec p\,^\prime)
               a(\vec p\,^\prime)a(\vec p\,)\rangle.
\label{pairs}
\eeq
Interferometry experiments measure the pair correlation function
\beq
   C(\vec q\,) = \frac{\left\{n_2(\vec p\,,\vec p\,^\prime)\right\}}
   {\left\{n(\vec p\,)n(\vec p\,^\prime)\right\}},
\label{corr}
\eeq
where the braces in the numerator denote an average over the total
momentum $\vec{P}$ for an ensemble of pairs from the same events at fixed
relative momentum $\vec q\, = (\vec p - \vec p\,^\prime)/2$, and in the
denominator they denote an average over particle pairs drawn from different
events.  Note that the 4-vector product $q \cdot P$ vanishes identically,
implying that the component of the relative momentum along the pair direction
$q_{\parallel}$ equals $(E_1-E_2)/2\beta$, where $\beta$ is the pair velocity
and $(E_1-E_2)/2$ is the relative pair energy.

    Since the Coulomb interaction conserves particles and the total momentum
of the pair, the final distribution $n_2(\vec p\,,\vec p\,^\prime)$ of pairs
of relative momenta $q$ is thus given in terms of the initial distribution of
pairs, $n_2^0(\vec p_0\,,\vec p_0\,^\prime)$, by
\beq
n_2(\vec p\,,\vec p\,^\prime)d^3q= n_2^0(\vec p_0\,,{\vec p_0\,}^\prime)d^3q_0
\label{transf}
\eeq
Equation (\ref{toy}), with changes in relative angles ignored, yields the
familiar Jacobian (see, e.g., \cite{gyu81}) $d^3q_0/d^3q = q_0/q$.  We can
assume to good accuracy that the Coulomb interactions between pairs of
particles negligibly affect the singles distributions in the denominator of
(\ref{corr}).  Then
\beq
  C(\vec q\,) = \frac{q_0}{q}C_0(\vec q_0\,)
  = \left(1 \mp \frac{2m_{red}e^2}{r_0 q^2}\right)^{1/2} C_0(\vec q_0\,).
\label{toydist}
\eeq

    To illustrate this toy model, we compare in Fig. 1 the predictions of Eq.
(\ref{toydist}) with E877 data for the $\pi^+\pi^-$, $\pi^-p$, and $\pi^+p$
systems produced in Au+Au collisions at the AGS \cite{e877}, assuming that the
bare correlation function $C_0$ equals unity.  Shown in this figure as dotted
lines are the results of the toy model for $r_0$ = 3 fm (rightmost curve), 9
fm, and 15 fm (leftmost curve), along with standard Gamow correction (solid
line).  Except at very small relative momenta $q\,\la\,10$ MeV/c, where
effects due to the finite momentum resolution of the experiment become visible
in the data, the model gives a good account of the data for $r_0$ in the range
of 9 - 15 fm.  By contrast, the Gamow factor considerably overpredicts the
data for all $q$ values shown here.  As we see, the raw correlation data for
non-identical particles contains information about the mean separation of
pairs when screening effects become negligible, summarized in the toy model by
the (possibly $q$ dependent) parameter $r_0$.  Then we show the Coulomb
correction factor deduced from Eq.  (\ref{toydist}) for $\pi^+\pi^+$ in Fig.
2a and $\pi^-\pi^-$ in Fig. 2b, for the same range of $r_0$ (as in Fig. 1, the
rightmost curve corresponds to $r_0$ = 3 fm).  Again we see that use of the
Gamow factor implies a correction which differs significantly from that of the
toy model.

    With the initial radius $r_0$ extracted from the unlike-sign data, one can
then construct the Coulomb correction for like-sign particles\footnote{It is,
in general, inadequate to correct the like-sign data with the inverse of the
unlike-sign Coulomb correction.  For $\pi-\pi$ correlations this approximation
is a rather good, except at very small $q$; it fails, however, for
correlations among heavier particles.}.  Dividing the ``raw'' E877 data by the
toy model correction factor, with $r_0 = 15 fm$, we obtain the correlation
function for like-sign pions (crosses) shown in Fig. 3a for $\pi^+\pi^+$ and
Fig. 3b for $\pi^-\pi^-$, which also show the correlation function (vertical
bars) derived by making the standard Gamow correction.  Using the Gamow factor
instead of the proper Coulomb correction leads to a correlation function which
is $\sim$ 30\% wider, implying a correspondingly reduced radius parameter.
Furthermore, the shape of the ``Gamow-corrected" correlation function has
considerable non-gaussian tails in the range $30 < q < 80 $ MeV/c.  These
tails do not exist in the raw correlation function and obscure the
interpretation of the data.

    We note also that the procedure described here for Coulomb corrections is
not restricted to one-dimensional correlation functions.  Since the Coulomb
correction depends only on the magnitude of the relative momentum of the pair
and not on its orientation, one should, for example, in an multi-dimensional
analysis in terms of $q_{out}$, $q_{side}$, and $q_{long}$, apply the
correction for, say, each bin of $q_{out}, q_{side}, q_{long}$ separately,
i.e., before projection onto the particular variable of interest.

\section{Connection with quantum-mechanical description}

    The physics of the toy model above is contained in the Coulomb wave
function describing the propagation of the pair.  To recall how this works we
suppress the total momentum of the pair, and focus only on the relative
momentum.  In the absence of Coulomb interactions the number of pairs of
relative momentum $\vec q\,$ is given by
\beq
        N_0(\vec q\,) = \int d^3r d^3r^\prime
          e^{i\vec q\,\cdot ({\vec r} - {\vec r}^\prime)}
     \langle J^\dagger(\vec r\,) J(\vec r\,^\prime)\rangle,
\label{JJ}
\eeq
where $J({\vec r}\,)$ is the amplitude for creating a pair at separation
$\vec r\,$, and the brackets denote an average over the event.  In the
presence of Coulomb interactions we have instead
\beq
N(\vec q\,) = \int d^3r d^3r^\prime \psi_C(\vec r\,) \psi_C^*(\vec r\,^\prime)
       \langle J^\dagger(\vec r\,) J(\vec r\,^\prime)\rangle,
\label{JJcoul}
\eeq
where $\psi_C(\vec r\,) = \psi_C(0) _1F_1(-i\eta;1;i(qr-{\vec q}\cdot
{\vec r}))$ is the relative Coulomb wave function for a pair of relative
momentum $\vec q\,$ at infinity.

    Pairs of low relative momentum have relatively low angular momentum, e.g.,
a pair produced at 10 fm separation with relative momentum 20 MeV/c can have
at most one unit of relative angular momentum.  Thus only the low partial wave
components of the Coulomb wave function enter Eq.  (\ref{JJcoul}) with
appreciable probability.  We consider here just s-waves, for which we employ
the WKB approximation to write the Coulomb wave function at radius $r$ outside
the classical turning point as\footnote{The WKB approximation for the s-wave
is quite good outside the collision volume for the parameters encountered in
HBT interferometry in ultrarelativistic collisions.  The condition for
validity of the approximation is $|\partial p(r)/\partial r| \ll p(r)^2$,
which for $r\ll a_0$, the region of interest, becomes the restriction, $r \ga
3/q^{3/2}a_0^{1/2}$.  For $\pi\pi$ (or $\pi$p) pairs with $q >$ 20 MeV/c, WKB
is reasonable for $r$ down to $\sim$ 5 fm (or $\sim$ 6 fm).}
\beq
\psi(r) \simeq \frac{1}{rp(r)^{1/2}q^{1/2}}\sin \phi(r),
\label{psiint}
\eeq
where the local relative momentum, measuring the rate of change of phase,
$\phi$, of the wave function, is given by
\beq
p(r) = \frac{d\phi}{dr} = \left(q^2 \mp \frac{2m_{red}e^2}{r}\right)^{1/2}.
\label{pr}
\eeq
(Equation (\ref{psiint}), with $\ell$-dependent $\phi(r)$ holds as well
for higher partial waves, $\ell > 0 $.)  The normalization of (\ref{psiint})
agrees with (\ref{psi0}) as $r\to 0$, while as $r\to\infty$, the Coulomb wave
function approaches
\beq
\psi(r) = \frac{1}{qr}\sin(qr -\eta\ln 2qr +\delta_0).
\label{psiinf}
\eeq

    The results of the toy model follow if we assume that the source
correlation function $\langle J^\dagger(\vec r\,) J(\vec r\,^\prime)\rangle$
is localized in both $r$ and $r^\prime$ around $r_0$. The correlation
function for s-waves (denoted by superscript $s$), the absence of Coulomb, is
\beq
    N_0^s(p) = \int d^3r d^3r^\prime
                \frac{\sin pr}{pr}\frac{\sin pr^\prime}{pr^\prime}
                 \langle J^\dagger(\vec r\,) J(\vec r\,^\prime)\rangle;
\label{JJ0s}
\eeq
then since in the region of any radius $r$ outside the turning point the
Coulomb wave function behaves locally as a free particle s-wave of momentum
$p(r)$, the correlation function is given by
\beq
  N^s(q) \approx \frac{p(r_0)}{q} N^s_0(p(r_0))
\label{N01}
\eeq
where the factor $p(r_0)/q$ arises from the denominators in Eq.
(\ref{psiint}) and (\ref{JJ0s}).  Consequently, $C(q) \simeq
C_0(p(r_0))p(r_0)/q$, the result in Eq.  (\ref{toydist}) with $q_0 = p(r_0)$.

    With the connection between the toy model and the Coulomb wave function
established we can now generalize the picture to extend to smaller values of
the source radius $r_0$.  In general, the effect of the Coulomb interactions
depends on the detailed structure of the source correlation function $\langle
J^\dagger(\vec r\,) J(\vec r\,^\prime)\rangle$; we expect this correlation
function to be approximately of the form
\beq
\langle J^\dagger(\vec r\,) J(\vec r\,^\prime)\rangle \approx
S(\vec r\, , \vec r\,^\prime) f({\vec r}\,^\prime - \vec r\,);
\label{source0}
\eeq
$S$, which defines the spatial region of the source, varies as a function
of its arguments on a scale of the size of the emitting region, and the
Fourier transform of $f$ defines the bare distribution of pairs, viz.,
\beq
f({\vec r}\,)=
\int \frac{d^3 q}{(2\pi)^3} e^{i\vec q\,\cdot \vec r\,} f(\vec q\,),
\label{baredistr}
\eeq
where, but for effects due to the finite size of the emitting region,
$N_0(\vec q) \sim f(\vec q\,)$.

    For a first orientation we can identify the bare correlation of Eq.
(\ref{baredistr}) with that experimentally observed for two-particle
correlations in $e^+e^-$ annihilation [15-18],
%
%
which indicate an HBT ``radius'' consistently below 1 fm.  Thus we expect
$f(\vec r - {\vec r\,}^\prime)$ to extend only over distances $\la$ 1 fm, or
equivalently the bare correlation to vary on momentum scales of several
hundred MeV/c.  Since typical source sizes from an analysis of pion
correlations following ultra-relativistic nucleus-nucleus collisions are of
the order of 5 - 10 fm \cite{e877,jacak}, it is a reasonable approximation to
neglect the size of the bare correlation and replace $f(\vec r - {\vec
r\,}^\prime)$ in Eq.  (\ref{source0}) by a delta function.  We write
\beq
\langle J^\dagger(\vec r\,) J(\vec r\,^\prime)\rangle \approx
\delta(\vec r\,-\vec r\,^\prime) N_0 S(r),
\label{source1}
\eeq
where $S(r)$, of unit strength, describes the distribution of
initial pair radii.  With (\ref{source1}) we find
\beq
C(q)/C_0(q) = N(q)/N_0 = \int d^3r |\psi_C(\vec r\,)|^2 S(r)
\label{NN0}
\eeq
(an equation tracing back to Ref. \cite{pratt86}).

    To illustrate the transition from the Gamow correction to the toy model we
take $S(r)$ to have a simple normalized gaussian form of range $r_0$:  $S(r)=
(2\pi)^{-3/2}r_0^{-3} \exp(-r^2/2r_0^2)$.  We show, in Fig. 4, for the
$\pi^+\pi^-$ system, the results of calculations of $C(q)/C_0(q)$ using
(\ref{NN0}) for $r_0$ = 1, 5, 9, and 18 fm (dash-dot curves, the highest for
$r_0$ = 1 fm, and falling with increasing $r_0$).  As $r_0\to 0$, the
projection of the square of the Coulomb wave function onto the source $S(r)$
converges to the standard Gamow correction (solid line).\footnote{For $r_0 <
0.1$ fm (not shown in Fig. 4) the difference between the Gamow correction and
a calculation with (\ref{NN0}) is less than 0.5\%.} For larger $r_0$ values it
rather quickly approaches the prediction of the toy model (shown here for an
initial radius of 9 fm as a dotted curve), implying that, for pairs
originating outside their classical turning point, the toy model provides an
adequate and reasonably accurate description of the Coulomb effects.

\section{The ``central" Coulomb potential}

    We next turn to the question of the effects of the Coulomb interactions of
the pair with the remaining particles.  This is a difficult many-body problem,
which we greatly simplify at this stage by assuming that the remaining
particles can be described by a central Coulomb potential, $Z_{eff}e^2/r$,
where in a central collison of nucleus A with nucleus B the effective charge
$Z_{eff}$ is of order of the total initial nuclear charge ($Z_A + Z_B$).  This
central potential accelerates positive mesons away and slows down the
negatives, effects described by the Coulomb wave functions for the potential.
The final momentum of any particle is related to the initial momentum $p_a$ at
production point $r_a$ by
\beq
\epsilon(p)= \epsilon(p_a) \pm \frac{Z_{eff}e^2}{r_a}.
\label{central}
\eeq
where $\epsilon(p) = (p^2+m^2)^{1/2}$.  (While Coulomb effects for the
relative momentum can be treated non-relativistically as in Eq. (\ref{toy}),
the individual momenta are generally relativistic.)  We ignore in this brief
discussion quantum mechanical suppressions or enhancements of the amplitude
for particle emission, as well as possible effects of angular changes in the
individual particle orbits on the particle distributions.  Then the single
particle distribution is modified by the central potential, analogously to Eq.
(\ref{transf}), by
\beq
n(\vec p\,) = n_0(\vec p_a\,)\frac{d^3p_a}{d^3p} =
\frac{p_a\epsilon(p_a)}{p\epsilon(p)} n_0(\vec p_a\,).
\label{singles1}
\eeq
Both the magnitude of the distribution as well as its argument are shifted.

    Although the central potential shifts the singles distribution, it cannot
introduce any correlations among emitted particles that have no initial
correlation in the absence of the central potential, e.g., for different
species or opposite charged pions, as one usually assumes.  If in the absence
of the central potential uncorrelated particles [$C(q)=1$] are emitted
in independent free particle states, then in the presence of the potential
they are emitted in Coulomb states for the central potential, but still
$n_2(\vec p\,,\vec p\,^\prime) = n(\vec p\,)n(\vec p\,^\prime)$ and $C(q)$
remains unity.

    For particles that are initially correlated as a consequence of
Bose-Einstein statistics, $n_2(\vec p\,,\vec p\,^\prime)$ and $n(\vec
p\,)n(\vec p\,^\prime)$ will be modified both by the Jacobians of the
transformations from initial to final momenta, and shifts of argument.
However, in forming $C(q)$, the effects of the Jacobians in the numerator and
denominator essentially cancel, and the primary effect is the shift in the
arguments:
\beq
   C(\vec q\,) = \frac{\left\{n_2(\vec p_a\,,\vec p_a\,^\prime)\right\}}
   {\left\{n(\vec p_a\,)n(\vec p_a\,^\prime)\right\}}.
\label{centcorr}
\eeq
Since positive particles are accelerated, the final momentum difference,
$\vec q = \vec p\, - \vec p\,^\prime$, of a positive pair will generally be
larger in magnitude than it is initially, while for negative pairs the final
momentum difference will generally be smaller.  Thus we expect the central
Coulomb potential to cause the size of collision volume extracted from
positive pairs to be smaller than the actual size, and that from negative
pairs larger than the actual size.  As an illustration consider a pair of
relativistic particles whose initial momenta $\vec p_a$ and $\vec
p_a\,^\prime$ are equal in magnitude to $p_a$, and final momenta $\vec p$ and
$\vec p\,^\prime$ equal in magnitude to $p$; then
\beq
   q = (p/p_a)q_a \simeq q_a \left(1 \pm \frac{Z_{eff}e^2/r_a}{p_a}\right).
\label{expand}
\eeq
where the upper sign refers to both particles positively charged and the
lower to both negatively charged.  For $Z\sim 150$, $r_a \sim 7$ fm and $p_a
\sim$ 300 MeV/c, the effect is an increase for positives (and a decrease for
negatives) in the observed scale of $C(q)$ and decrease (or increase) in the
extracted radius of ten percent.

    GB is grateful to S. Koonin for his classical wisdom, to J. Popp for
helpful calculations, including those shown in Fig. 4, and T. Humanic and H.
Str\"obele for discussions.  This work was supported by NSF Grant PHY94-21309.

\bibliographystyle{unsrt}

\section{Figure captions}

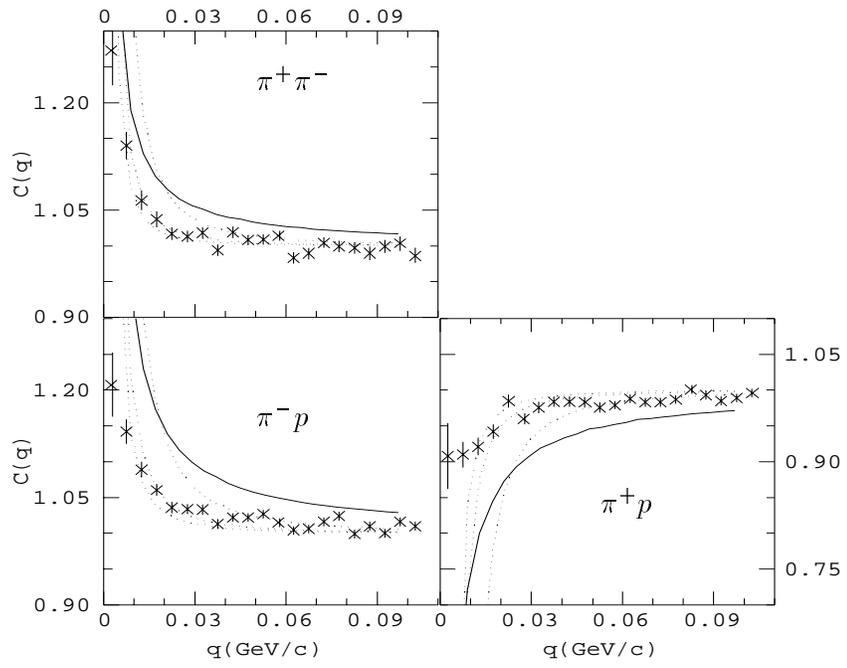
    FIG. 1. Comparison of the toy model, Eq.  (\ref{toydist}) for $r_0$ = 3 fm
(rightmost curve), 9 fm, and 15 fm (leftmost curve), with E877 data
\cite{e877} for the systems $\pi^+\pi^-$, $\pi^-p$, and $\pi^+p$, assuming a
bare correlation function $C_0 = 1$.  Solid line:  Gamow correction.\\

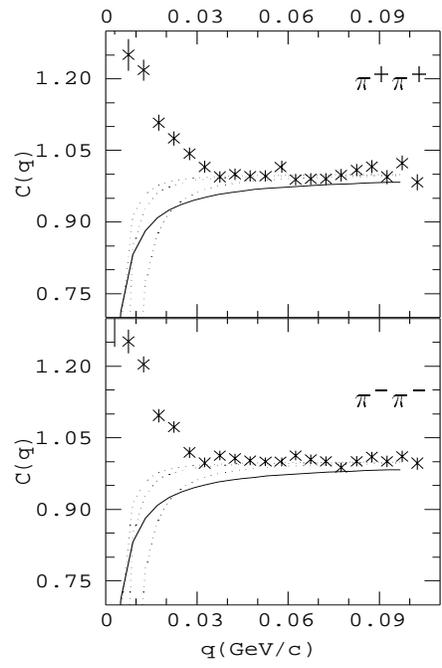
    \noindent FIG. 2. Coulomb correction, Eq.  (\ref{toydist}) for the systems
$\pi^+\pi^+$ and $\pi^-\pi^-$ for the same range of $r_0$ as in Fig. 1
(the rightmost curve corresponds to $r_0$ = 3 fm).\\

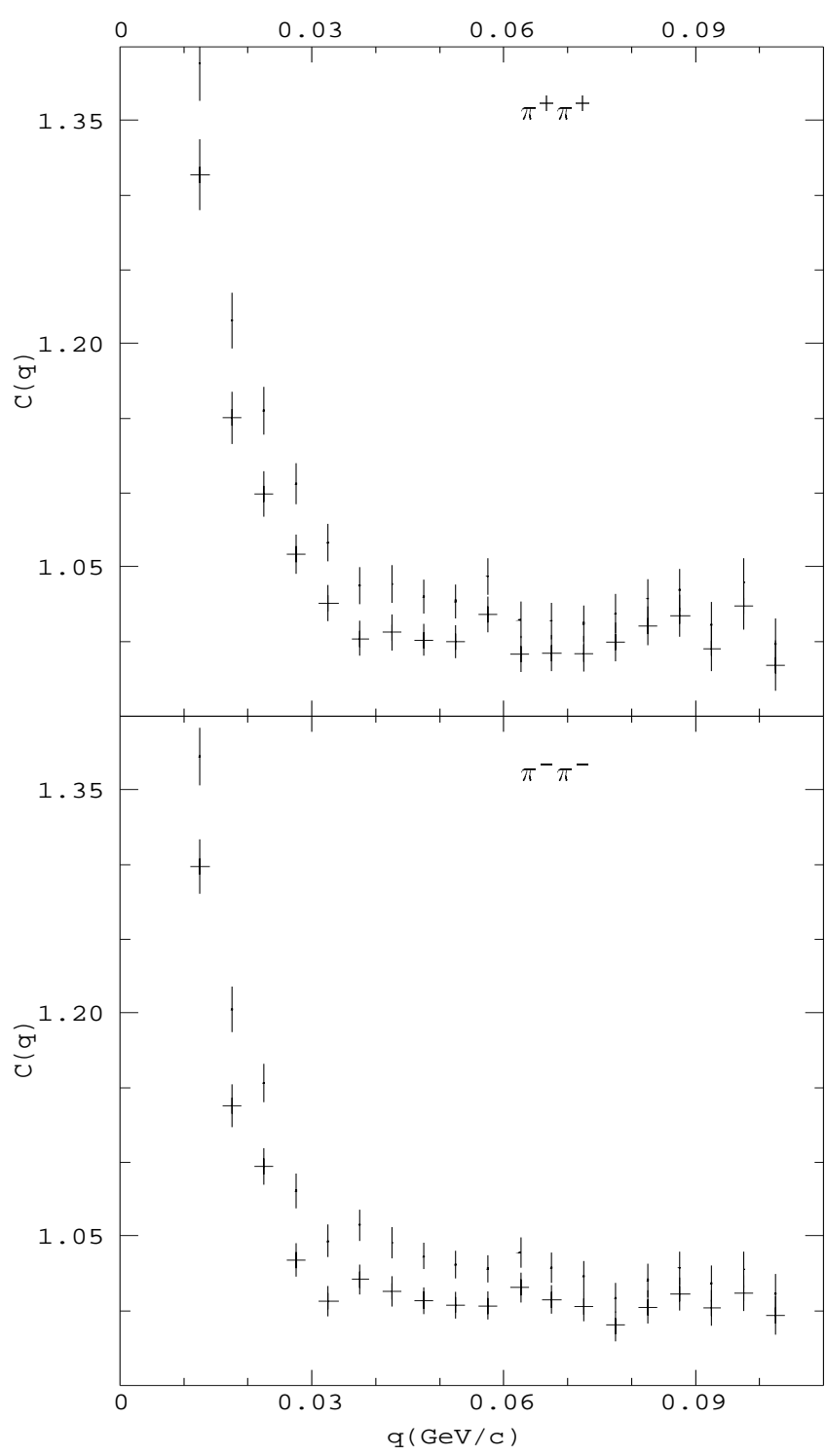
    \noindent FIG. 3. Toy model calculation of $C(q)$ for like-sign pions
(crosses), compared with the correlation function derived by making the
standard Gamow correction (vertical bars).\\

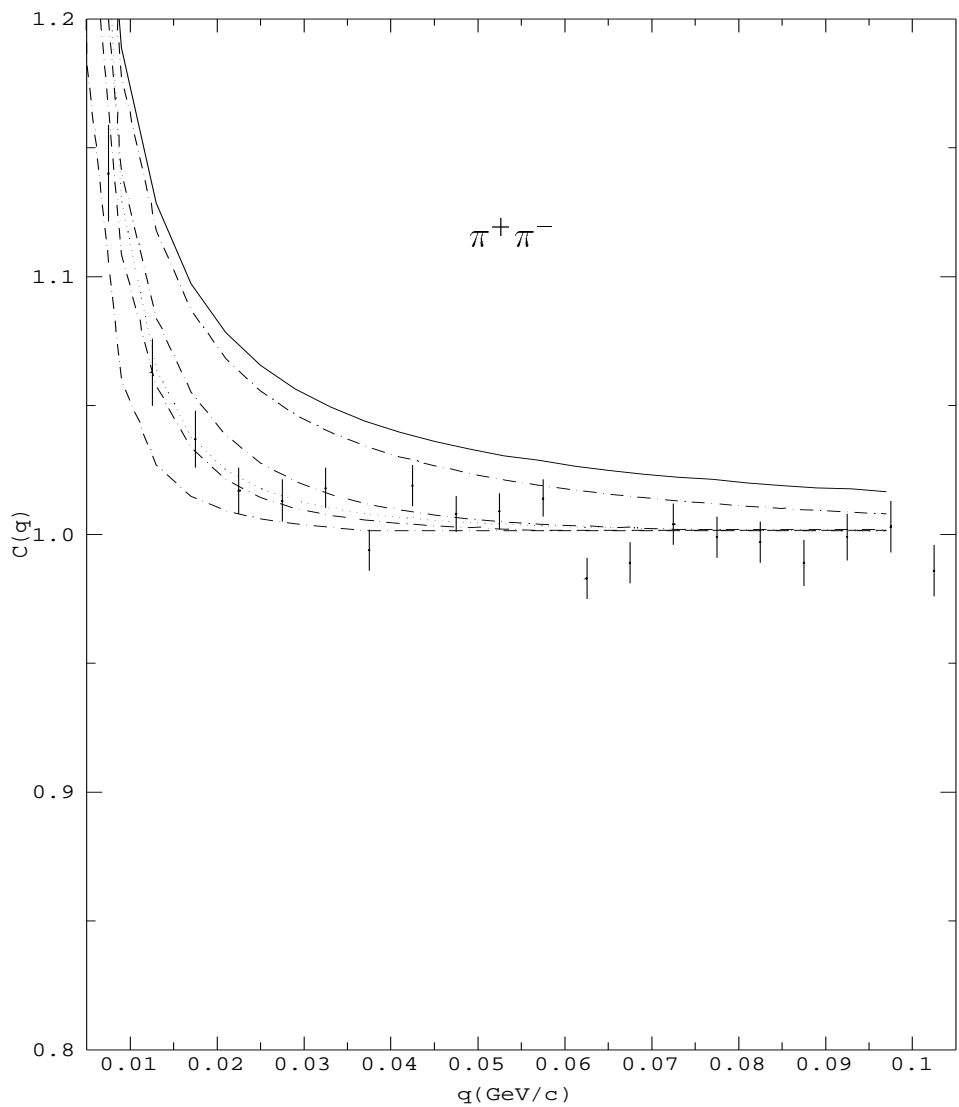
    \noindent FIG. 4. Transition from the toy model (dotted line, with $r_0$ =
9 fm) to the Gamow correction (solid line) with decreasing source size,
calculated from Eq.  (\ref{NN0}) (dash-dot curves).  From highest to lowest
dash-dot curves the source range $r_0$ is 1, 5, 9, 18 fm.  The data shown are
for $\pi^+\pi^-$.

\end{document}